\documentclass{aa}  
\usepackage{graphicx}
\usepackage{txfonts}
\usepackage{xcolor}
\usepackage[nolist]{acronym}
\usepackage{makecell}
\usepackage{booktabs}
\usepackage{comment}
\usepackage{anyfontsize}
\usepackage{footmisc}
\usepackage[colorlinks=true,linkcolor=blue]{hyperref}
\hypersetup{citecolor=blue, urlcolor=blue}
\usepackage{orcidlink}
\usepackage{siunitx}
\usepackage{amsmath}
\usepackage{nccmath}
\usepackage{tablefootnote}
\usepackage{soul}

\newcommand{\askap}{ASKAP J1118--5437}
\newcommand{\g}{G289.6+5.8}
\newcommand{\gaia}{DR3~5346688272924690432}

\newcommand{\igr}{IGR~J11187--5438}
\newcommand{\mass}{2MASS~J11182121--5437286}
\newcommand{\erass}{1eRASS~J111821.2--543729}

\begin{acronym}
\acro{dm}[DM]{dispersion measure}
\acro{lmxb}[LMXB]{low-mass X-ray binary}
\acro{pwn}[PWN]{ pulsar wind nebula}
\acro{rfi}[RFI]{radio frequency interference}
\end{acronym}

\begin{document} 

\titlerunning{A newborn spider system at the centre of a radio shell}\authorrunning{S. Lazarević et al.}


\title{A newborn spider system at the core of a radio shell: \\Evidence for a low-energy supernova}
   \author{S.\,Lazarevi\'c\inst{1,2,3}\thanks{Corresponding\,authors:\,\href{mailto:s.lazarevic@westernsydney.edu.au}{s.lazarevic@westernsydney.edu.au},\\\href{robert.brose@desy.de}{robert.brose@desy.de}}\orcidlink{0000-0001-6109-8548}
          \and
          R.\,Brose\inst{4}\footnotemark[1]\orcidlink{0000-0002-8312-6930}
          \and
          L.\,M.\,Oskinova\inst{4}\orcidlink{0000-0003-0708-4414}
          \and
          M.\,Chernyakova\inst{5,6}\orcidlink{0000-0002-9735-3608}
          \and
          S.\,Dai\inst{2}\orcidlink{0000-0002-9618-2499}
          \and
          O.\,Kargaltsev\inst{7}\orcidlink{0000-0002-6447-4251}
          \and
          S.\,Freund\inst{8}\orcidlink{0009-0006-3658-4935}
          \and
          C.\,Maitra\inst{9,8}\orcidlink{0000-0002-0766-7313} 
          \and
          M.\,D.\,Filipovi\'c\inst{1}\orcidlink{0000-0002-4990-9288}
          \and
          P.\,G. Edwards\inst{2}\orcidlink{0000-0002-8186-4753} 
          \and 
          I.\,El\,Mellah \inst{10}\orcidlink{0000-0003-1075-0326}
          \and
          Z.\,Guo\inst{11,12}\orcidlink{0000-0003-0292-4832}
          \and
          J.\,Osses\inst{11,12}\orcidlink{0009-0007-6769-7075}
          \and
          B.\,van\,Soelen\inst{13}\orcidlink{0000-0003-1873-7855}
          \and
          S.\,B.\,Potter\inst{14,15}\orcidlink{0000-0002-5956-2249}
          \and
          R.\,Kothes\inst{16}\orcidlink{0000-0001-5953-0100}
          \and
          G.\,P.\,Rowell\inst{17}\orcidlink{0000-0002-9516-1581}
          \and
          V.\,Velovi\'c\inst{1}\orcidlink{0000-0002-0416-3267}
          \and
          A. Ahmad\inst{1}\orcidlink{0000-0002-0457-3661}
          \and
          B.\,D.\,Ball\inst{18}\orcidlink{0009-0003-2088-9433} 
          \and
          C.\,Burger-Scheidlin\inst{6}\orcidlink{0000-0002-7239-2248}
          \and
          T.\,J.\,Galvin\inst{19}\orcidlink{0000-0002-2801-766X}
          \and
          Y.\,A.\,Gordon\inst{20}\orcidlink{0000-0003-1432-253X}
          \and
          A.\,M.\,Hopkins\inst{21}\orcidlink{0000-0002-6097-2747}
          \and
          D.~Leahy\inst{22}\orcidlink{0000-0002-4814-958X}
          \and
          J.\,Pritchard\inst{2}\orcidlink{0000-0003-1575-5249}
          \and
          J.\,West\inst{16,23}\orcidlink{0000-0001-7722-8458}
          }

   \institute{Western Sydney University, Locked Bag 1797, Penrith South DC, NSW 2751, Australia 
   \and 
   Australia Telescope National Facility, CSIRO, Space and Astronomy, PO Box 76, Epping, NSW 1710, Australia
   \and 
   Astronomical Observatory, Volgina 7, 11060 Belgrade, Serbia
   \and 
   Institute of Physics and Astronomy, University of Potsdam, 14476 Potsdam-Golm, Germany 
   \and 
   School of Physical Sciences and Centre for Astrophysics \& Relativity, Dublin City University, Glasnevin, Dublin, D09 W6Y4, Ireland
   \and 
   Astronomy \& Astrophysics Section, School of Cosmic Physics, Dublin Institute for Advanced Studies, DIAS Dunsink Observatory, Dublin D15 XR2R, Ireland
   \and 
   Department of Physics, The George Washington University, 725 21st St. NW, Washington, DC 20052, USA
   \and 
   Max-Planck-Institut für extraterrestrische Physik, Gießenbachstraße 1, D-85748 Garching bei München, Germany
   \and 
   Inter University Centre for Astronomy\& Astrophysics, Ganesh-khind, Pune 411007, India
   \and 
   Departament de Física, EEBE, Universitat Politècnica de Catalunya, c/Eduard Maristany 16, 08019 Barcelona, Spain
   \and 
   Instituto de Física y Astronomía, Universidad de Valparaíso, ave. Gran Bretaña, 1111, Casilla 5030, Valparaíso, Chile 
   \and 
   Millennium Institute of Astrophysics, Nuncio Monseñor Sotero Sanz 100, Of. 104, Providencia, Santiago, Chile
   \and 
   Department of Physics, University of the Free State, PO Box 339, Bloemfontein 9300, South Africa
   \and 
   South African Astronomical Observatory, PO Box 9, Observatory 7935 Cape Town, South Africa
   \and 
   Department of Physics, University of Johannesburg, PO Box 524, Auckland Park 2006, South Africa
   \and 
   Dominion Radio Astrophysical Observatory, Herzberg Astronomy \& Astrophysics, National Research Council Canada, P.O. Box 248, Penticton, BC V2A 6J9, Canada
   \and 
   School of Physics, Chemistry and Earth Sciences, The University of Adelaide, Adelaide, 5005, Australia
   \and 
   Department of Physics, University of Alberta, Edmonton, Alberta, T6G 2E1, Canada
   \and 
   Australia Telescope National Facility, CSIRO, Space and Astronomy, PO Box 1130, Bentley, WA 6151, Australia
   \and 
   Department of Physics, University of Wisconsin-Madison, 1150 University Ave., Madison, WI 53706, USA
   \and 
   School of Mathematical and Physical Sciences, Macquarie University, 12 Wally’s Walk, Macquarie Park, 2109, NSW, Australia
   \and 
   Department of Physics and Astronomy, University of Calgary, Calgary, Alberta, T2N 1N4, Canada
   \and 
   School of Natural Sciences, University of Tasmania, PO Box 807, Sandy Bay, TAS 7006 Australia
   }

\date{Received Month DD, YYYY; accepted Month DD, YYYY}

\abstract{
In a search for low surface brightness radio nebulae using the ASKAP--EMU survey, we discovered a faint radio shell, \g, and its central point radio source at the position of the soft $\gamma$-ray source \igr. 
The central radio source is spatially coincident with a previously known low-mass X-ray binary~(LMXB) with an M-type donor star. However, the newly determined Gaia DR3 distance of 267\,pc and correspondingly low X-ray luminosity~($3\times10^{31}\,$erg\,s$^{-1}$) cast doubt on the LMXB classification. Neither radio nor X-ray pulsations are detected. Chance-alignments between radio shell, central radio source, optical star, $\gamma$-ray, and X-ray sources appear unlikely. By combining all currently available evidence, we conclude that \g\ is a remnant of a low-energy core-collapse explosion of an intermediate mass star~($\sim 8M_\odot$) in a binary system with an M-type secondary, which remained bound after the explosion. In this scenario, \g\ is a supernova remnant, while the central $\gamma$- and X-ray source is associated with a young neutron star driving a pulsar wind interacting with its M-type stellar companion, making \igr\ a nascent spider-type X-ray binary.}


\maketitle

\begin{figure*}[ht!]
\centering
\includegraphics[trim=0 0 0 0, width=0.75\linewidth]{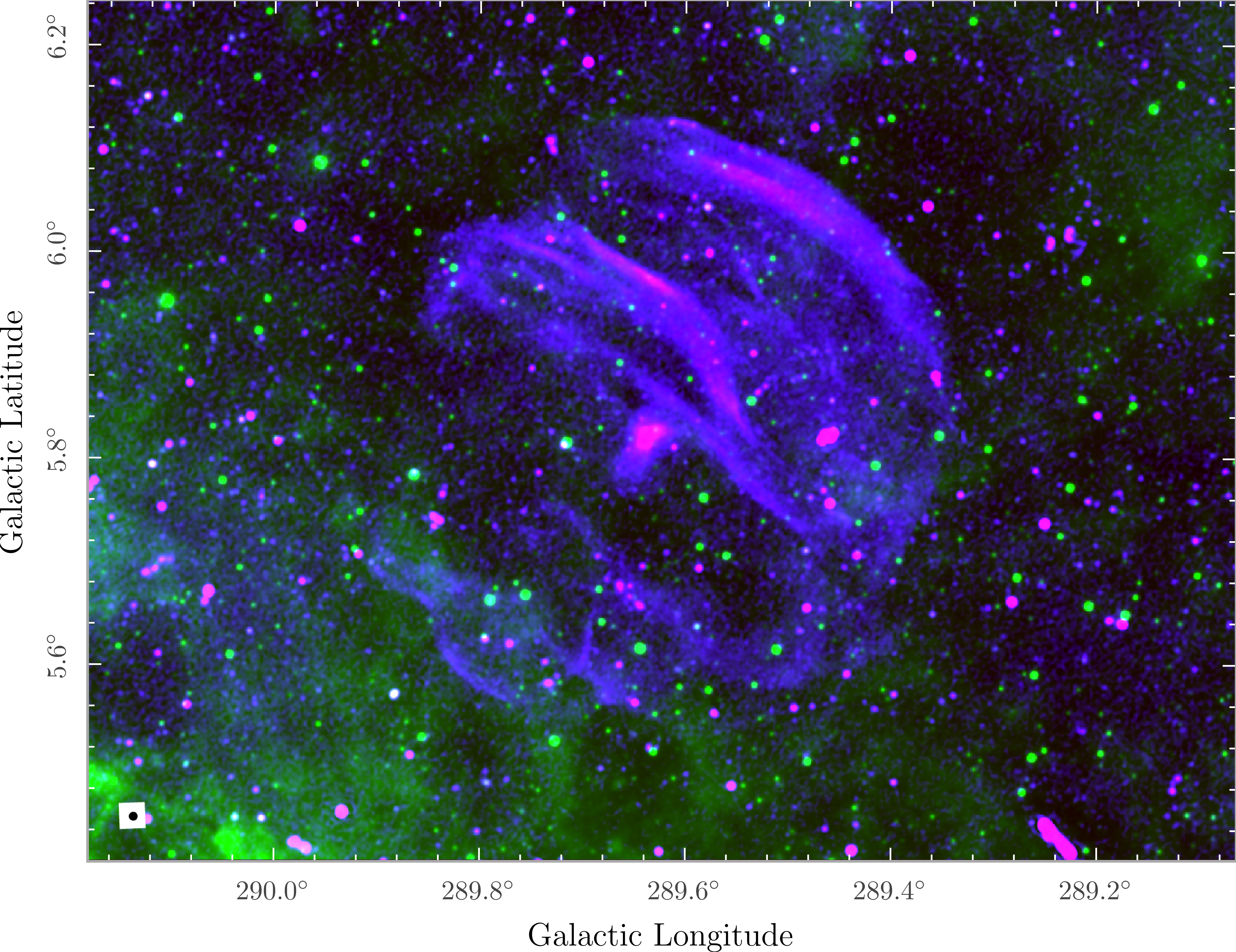}
\caption{RGB composite image of \g\ showing the large-scale radio structure of the remnant. The image combines the ASKAP--EMU 944\,MHz total intensity map~(red and blue) with WISE 12\,$\mu$m emission~(green). To emphasise the structure of the source, the radio image is displayed in two different scalings: the red layer uses a linear stretch to highlight the bright filaments and central region, and the blue layer a logarithmic stretch to enhance the low surface brightness emission. The 15\arcsec\ synthesised circular beam is shown in the lower-left corner. The $\gamma$-ray source \igr\ lies within the compact nebula at the centre.}
\label{fig:rgb}
\end{figure*}

\section{Introduction}

High-cadence optical surveys over the past decade have revealed a growing population of transients with luminosities intermediate between classical novae and supernovae~(SNe). These {\em gap} transients originate from diverse phenomena, including pre-SN outbursts, SN impostors, stellar mergers, and faint core-collapse SNe \citep[ccSNe;][for a review]{2019NatAs...3..676P}. A subset of {\em gap} transients is intermediate-luminosity red transients~(ILRTs), characterised by a slowly rising light curve, a linear decline over $\sim$$4$ months, and spectral evolution from blue to progressively redder colours. While heavily dust-obscured progenitors have been observed for some ILRTs, no post-explosion remnants have been detected \citep[e.g.,][]{2009ApJ...695L.154B}. This absence suggests a catastrophic origin, with low-energy electron-capture supernovae~(ecSNe) proposed as an underlying mechanism \citep{2019NatAs...3..676P}.

Theoretically, ecSNe are expected to occur from stars of initial mass 8--10\,M$_\odot$ that develop degenerate O–Ne–Mg cores undergoing rapid electron capture \citep[][and references therein]{1987ApJ...322..206N}. These explosions are predicted to be under-energetic compared to typical ccSNe, with energies of $\sim$$10^{50}\,$erg \citep{2006A&A...450..345K}, or even as low as $10^{48}\,$erg based on observational constraints \citep{2020A&A...639A.103S}. Alternative scenarios propose even weaker explosions~($\sim$$10^{47}\,$erg), driven by neutrino mass loss during core collapse \citep{2013ApJ...769..109L}. From an evolutionary perspective, ecSNe are favoured in binary systems. Owing to their low explosion energies and lower natal kicks, binaries are more likely to survive such events. In fact, binaries that survive an ecSNe are considered a key formation channel for \acp{lmxb}, in which a neutron star~(NS) or black hole accretes from a low-mass companion~\citep{Podsiadlowski2004, Ivanova2008}. 

During an ecSN, ejecta with kinetic energies 10$^{47}$--10$^{50}$\,erg expand supersonically into the interstellar medium, driving shocks capable of accelerating relativistic electrons and producing synchrotron radio emission. Given that the observed flux scales as $F\propto\dot{E}/D^2$, nearby ecSN remnants may resemble more energetic supernova remnants~(SNRs) at much larger distances, making reliable distance estimates essential for their identification. Although ecSNe are theoretically expected to occur at $\leq5$\% of the ccSN rate \citep[][]{2009ApJ...705.1364T, 2009MNRAS.398.1041B}, no Galactic remnant of an ecSN has yet been conclusively identified. 

New generation radio facilities such as ASKAP \citep{2021PASA...38....9H} and MeerKAT~\citep{meerKAT} provide the sensitivity required to detect such faint Galactic remnants. Their deep surveys have nearly doubled the number of known and candidate SNRs to $\sim$$600$, most with low surface brightness~\citep{2023MNRAS.524.1396B,2025ApJ...988...75B,2025A&A...693A.247A}. Based on estimated ecSN rates, up to $\sim$24 of these remnants could originate from low-energy explosions, although radio data alone are insufficient for identifying such events. 

The paper is structured as follows. In Sect.~\ref{sec:radio}, we report the discovery of a radio shell \g\ and its central source. In Sect.~\ref{sec:central}, we analyse the $\gamma$-ray and X-ray source \igr\ located at the centre of the shell. In Sect.~\ref{sec:discussion}, we argue that a common ILRT origin for both the central source and the radio remnant provides the most plausible explanation of the multiwavelength observations. Our conclusions are given in Sect.~\ref{sec:conclusion}.

\begin{figure*}[ht]
 \centering
\includegraphics[trim=0 0 15 0, width=\columnwidth]{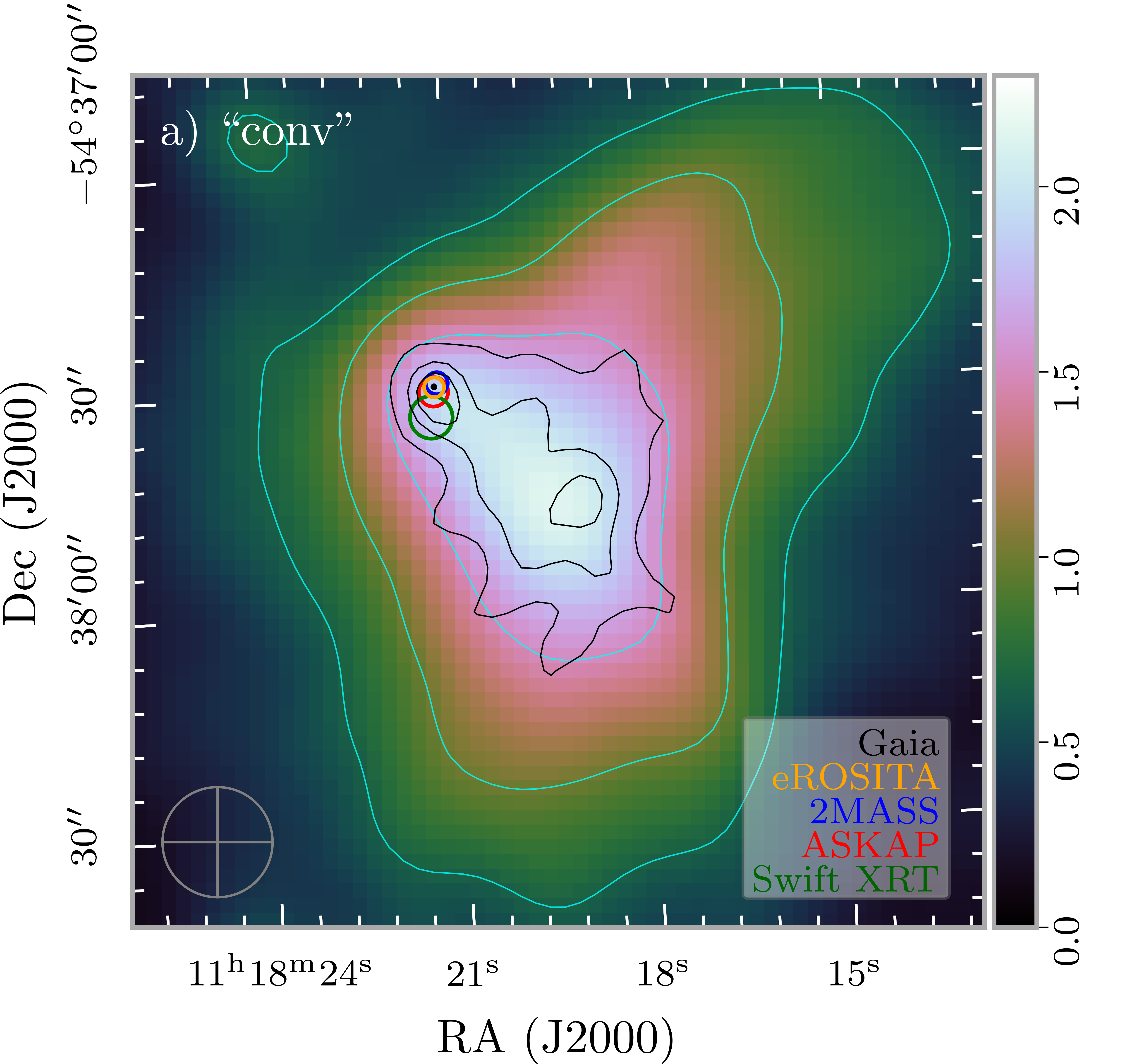}
\includegraphics[trim=15 0 0 0, width=\columnwidth]{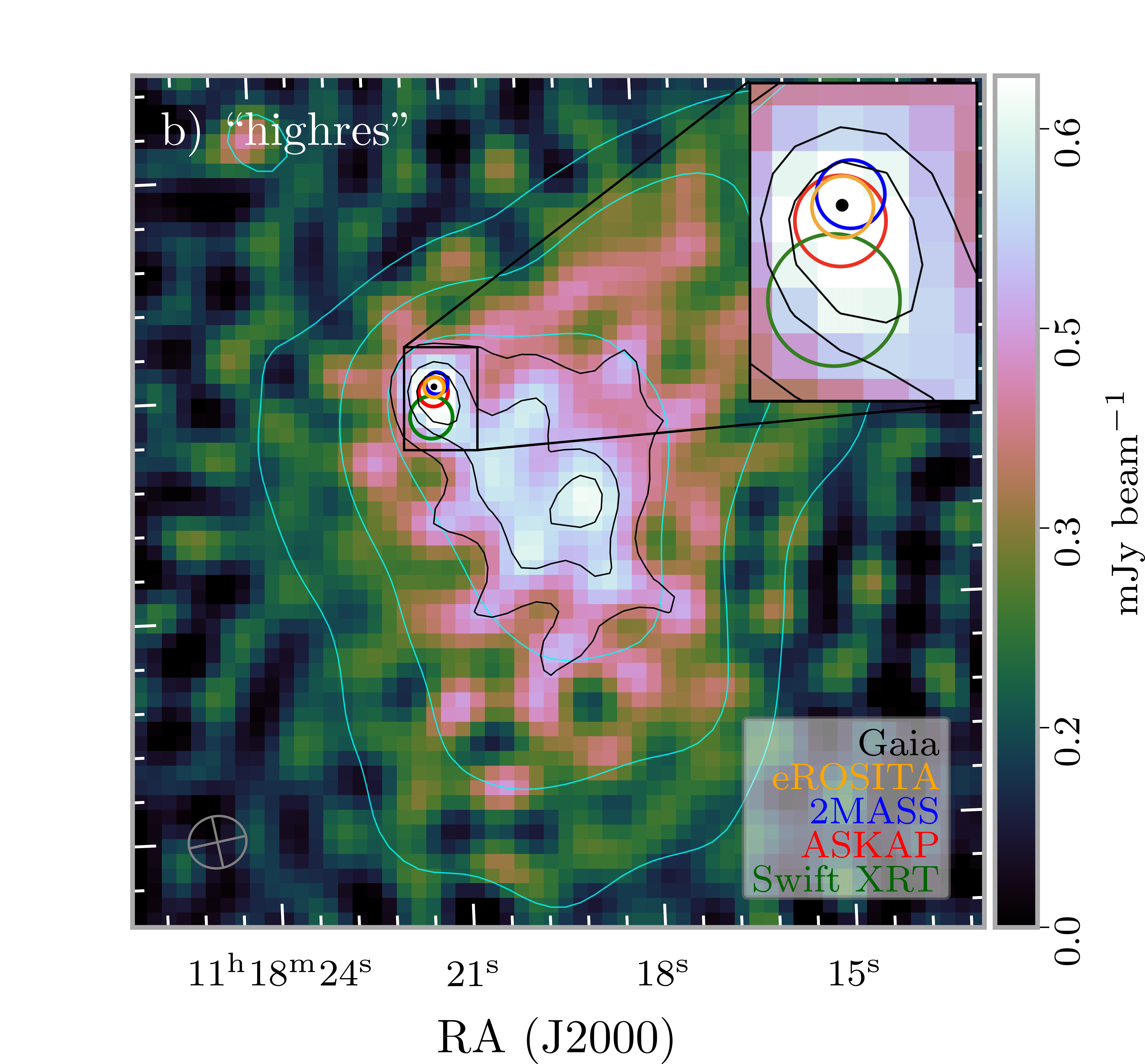}
\caption{ASKAP--EMU view of the central region of \g. Panel (a) shows the standard ``conv'' image, and panel (b) the higher-resolution ``highres'' image.  The ``conv'' image has a uniform resolution of 15$^{\prime\prime}\times$15$^{\prime\prime}$, while the ``highres'' image has a resolution of 7.9$^{\prime\prime}\times$7.1$^{\prime\prime}$. The synthesised beam sizes are given in the lower-left corners. Cyan contours correspond to the ``conv'' image at levels of 0.625, 1 and 1.5 mJy\,beam$^{-1}$. Black contours trace the ``highres'' image at 0.4, 0.5 and 0.55 mJy\,beam$^{-1}$ levels. 
Coloured circles mark the sources listed in Table~\ref{tab:associations}, with circle radii indicating their positional uncertainties. The inset in panel (b) presents a zoomed view of the boxed region surrounding these sources. 
}
\label{fig:conv-highres}
\end{figure*}

\section{Low surface brightness radio-shell: \g}
\label{sec:radio}

The strikingly beautiful extended radio source \g~(see Fig.\,\ref{fig:rgb}) was serendipitously discovered in ASKAP observations 
obtained as part of the Evolutionary Map of the Universe survey~\citep[EMU;][]{2021PASA...38...46N,2025PASA...42...71H}. 
The source is roughly circular, with an angular extent of $\sim$$36.6^\prime\times34.4^\prime$ and a centre at~(R.A.,\,Dec)$_\text{J2000}$ = 11:18:15.1, --54:37:38. Its morphology shows an asymmetric brightness distribution with two bright filaments in the central-to-northwest region embedded in faint diffuse emission, and a weaker shell along the south. A bright compact nebula dominates the central region. A zoomed-in view of the centre~(Fig.\,\ref{fig:conv-highres}) reveals a point source ASKAP J1118--5437~(Table\,\ref{tab:associations}). Details of the ASKAP observations and EMU data products are provided in Appendices~\ref{sec:askap-obs} and \ref{sec:emu-data}. 

To investigate the nature of the radio emission, we compare the ASKAP--EMU 944~MHz image with the WISE 12~$\mu$m map of \g~\citep{2010AJ....140.1868W} shown in Fig.\,\ref{fig:rgb}. 
Such radio–infrared~(IR) comparisons are commonly used to distinguish non-thermal from thermal emission, as thermal sources typically show correlated IR emission from heated dust \citep{2023MNRAS.524.1396B,2025A&A...693A.247A}. The absence of an IR counterpart suggests a non-thermal origin for the radio emission, characteristic of SNRs.

Following the method described in~\cite{2023MNRAS.524.1396B,2025ApJ...988...75B}, we measure integrated flux densities of $1.42\,\pm\,0.24$\,Jy for the entire \g\ and $0.03\,\pm\,0.01$\,Jy for its central nebula. Because the extended diffuse emission of \g\ has low surface brightness and may be partially resolved out in the interferometric observations~\citep{2025PASA...42...71H}, we restricted the spectral analysis to the central region and the brightest filamentary structures. The spectral indices were derived using two independent approaches: in-band spectral fitting and a Taylor-term analysis of the ASKAP data. The full methodology is described in Appendix~\ref{sec:spectral}.

For the central region, we measured spectral indices within three contour-defined areas~(cyan contours in Fig.\,\ref{fig:conv-highres}). In-band imaging yields values of $\alpha$\,=\,$-$0.51\,$\pm$\,0.26,\,$-$0.38\,$\pm$\,0.18 and $-$0.28\,$\pm$\,0.10 for 
the outer, middle, and inner regions, respectively. These results are consistent with the spectral indices obtained using the Taylor-term technique~(Table \ref{tab:flux_central}). The spectral index flattening toward the core indicates the presence of freshly accelerated high-energy electrons, indicating the presence of ongoing particle acceleration~\citep{2017ASSL..446....1K}. For the brightest filaments, we derive spectral indices of $\alpha$\,=\,$-0.76\pm0.56$ for the outer filament and $-0.56\pm0.32$ for the inner filament, again consistent with the Taylor-term analysis. Although the uncertainties are relatively large, the measured values are broadly consistent with the canonical spectral index expected for non-thermal synchrotron emission~\citep{book2}.

\section{The central source of the radio nebula: CSRN}
\label{sec:central}

\begin{table*}[htb!]
\caption{Multiwavelength counterparts of CSRN~(see accompanying Fig.\,\ref{fig:conv-highres}). Pos. error lists the quoted 1$\sigma$ statistical positional uncertainty reported by the respective catalogues, and Separation lists the angular distance between each source and Gaia\,32~(Gaia~\gaia).}
\label{tab:associations}
\begin{tabular}{ | c | l | l l | c | c |l|} 
\hline
\makecell{} & \makecell{Catalog ID} & \makecell{R.A.} & \makecell{Dec.} & \makecell{Pos.\,error \\ $[$arcsec$]$} & \makecell{Separation \\ $[$arcsec$]$} & \makecell{Reference}\\  
\hline
\hline
1 & \askap & 11:18:21.27 & --54:37:29.83 & 2 &  0.72 & This work \\ 
2 & \igr\ & 11:18:21.120 & --54:37:33.60 & 207.6  & 4.63 & \small\cite{Bird2016} \\
3 & 2SXPS J111821.3--543733	& 11:18:21.32 & --54:37:33.28 & 2.9 & 4.2 & \small\cite{2020ApJS..247...54E}\\
4 & \mass\ & 11:18:21.211 & --54:37:28.63 &  1.5 &  0.54 & \small\cite{2003tmc..book.....C} \\  
5 & \erass\ & 11:18:21.254 & --54:37:29.19 & 1.35   & 0.07 & \small\cite{2024AA...682A..34M} \\ 
6 & Gaia \gaia\ & 11:18:21.257 & --54:37:29.12 & $10^{-4}$ &  & \small\cite{2023AA...674A...1G} \\ 
\hline
\end{tabular}
\end{table*}

Long before \g\ was discovered, a soft $\gamma$-ray source \igr\ was detected by the {\em INTEGRAL}\,IBIS/ISGRI telescope~\citep{2007ApJS..170..175B, Bird2016}. Follow-up observations with the {\em Swift}\,XRT~\citep{2008AA...482..731R} confirmed the X-ray counterpart and constrained the line-of-sight absorption, deriving a hydrogen column density of $N_{\rm H}\,\approx\,2.8\times\,10^{21}$~cm$^{-2}$. \cite{2013A&A...560A.108C} identified potential infrared and optical counterparts in the 2MASS and the DSS\,II catalogues, based on which \igr\ was classified as a LMXB. However, the unknown distance to the source precluded a reliable determination of its X-ray luminosity.

Our discovery of a radio nebula \g, encompassing IGR J11187--5438, reignites interest in the true nature of the source. For convenience, we refer to this source as the central source of the radio nebula~(CSRN) throughout the rest of the paper.  

Crucial insights into CSRN come from the {\em eROSITA} point source catalogue \citep{2024AA...682A..34M}, which provides the most precise localisation among the available high-energy observations of this region. The X-ray point source \erass\ positionally coincides with both the {\em INTEGRAL} detection and the radio point source~(see Table\,\ref{tab:associations}), and no other X-ray sources are detected in its immediate vicinity\footnote{The nearest X-ray source in the eRASS1 catalogue lies $\sim$2$'$ from \erass, while the closest neighbour of the 2SXPS J111821.3--543733 in the {\em Swift} catalogue is 3.2$'$ away.}. 
The combined spectra of  {\em eROSITA}, {\em Swift}\,XRT, {\em Swift}\,BAT and {\em INTEGRAL} align well~(see Fig.\,\ref{fig:xrt-bat}), indicating a common origin for all four datasets and yielding a relatively hard photon index of $\Gamma$\,=\,1.36\,$\pm$\,0.03. The details of the spectral extraction and analysis are Sect.~\ref{sec:x-ray}. 

In the eRASS1 coronal source catalogue \citep{2024A&A...684A.121F}, our source is associated with Gaia \gaia, hereafter Gaia\,32, with a matching probability of 92\%. Gaia\,32 is an M3-type star located at a distance of $d$\,$\approx $\,267$\pm$\,9~pc. However, the observed X-ray properties are inconsistent with those expected from typical coronal emission of late-type stars. The source exhibits an unusually hard X-ray spectrum~($\Gamma$\,=\,1.36\,$\pm$\,0.03) and an extreme fractional X-ray flux of $\log(F_X/F_\mathrm{bol})$\,=\,$-$0.7, far above the levels expected for stellar coronal emission. Explaining such emission through coronal activity alone would require an exceptionally strong flare, yet the eRASS1 light curve is relatively stable and shows no evidence for major flaring activity. Therefore, \citet{2024A&A...684A.121F} classify the source as non-coronal, implying that the X-ray emission associated with Gaia\,32 is unlikely to originate solely from coronal activity~(see 
Sect.~\ref{sec:coronal} for catalogue details). 

We obtained optical spectroscopy of Gaia 32 with the 4.1~m SOAR telescope. 
The spectrum shows features of a cool star consistent with an M3 type~(Fig.\,\ref{fig:optical}). However, contamination from a bright star located only $6''$ away~(Fig.\,\ref{fig:ir}) reduces the signal-to-noise ratio 
and may obscure subtle signatures of interaction with a compact companion. 
We also obtained optical photometry with the 1.9~m and 1.0~m SAAO telescopes, which shows no significant variability. Details of the optical observations and data reduction are presented in Appendix~\ref{sec:optical}. Additionally, Gaia astrometry reveals no notable \textit{astrometric\_excess\_noise} compared to nearby stars of similar brightness, providing no evidence for binarity.


Finally, we quantify the probability of a chance alignment between the radio shell, the X-ray detections, and the Gaia/2MASS counterpart. The resulting probability is $p\lesssim10^{-3}$, indicating that the observed configuration is unlikely to occur by chance and strongly supports a physical association between these sources~(see Sect.~\ref{sec:ChanceAlignment} for a detailed probability calculation).

\subsection{X-ray properties of CSRN}
\label{sec:x-ray}

To investigate the nature of the CSRN, we examine the positional consistency of the multiwavelength counterparts and the X-ray properties of available X-ray observations. In Table~\ref{tab:associations}, we list all CSNR counterparts together with their positional uncertainties and angular separations relative to Gaia\,32. Their positions and uncertainties are overlaid in Fig.\,\ref{fig:conv-highres}a. 

The ASKAP point source and \erass\ are both consistent with the Gaia\,32 position within their respective uncertainties. The position of 2SXPS J111821.3--543733 from the 2SXPS Catalogue of Swift XRT Point Sources~(clean version) is offset by approximately 4\arcsec\ from \erass, which has the smallest X-ray positional uncertainty. Given the {\em Swift\,XRT} statistical uncertainty of 2.9\arcsec~(90\% confidence), this offset corresponds to only a $\sim$$2\sigma$ deviation and is therefore not significant. Moreover, the {\em Swift\,XRT} pointing accuracy carries an additional $\sim$3\arcsec\ systematic uncertainty that is not accounted for here. Although the INTEGRAL IBIS–ISGRI detection is spatially associated with \erass, its large positional uncertainty~($\approx$3.5\arcmin) prevents a meaningful astrometric comparison. The association is instead supported by the absence of any alternative X-ray sources within the INTEGRAL localisation region.  

\begin{figure}[htp!]
\includegraphics[trim=10 10 0 0, width=0.71\columnwidth, angle=-90]{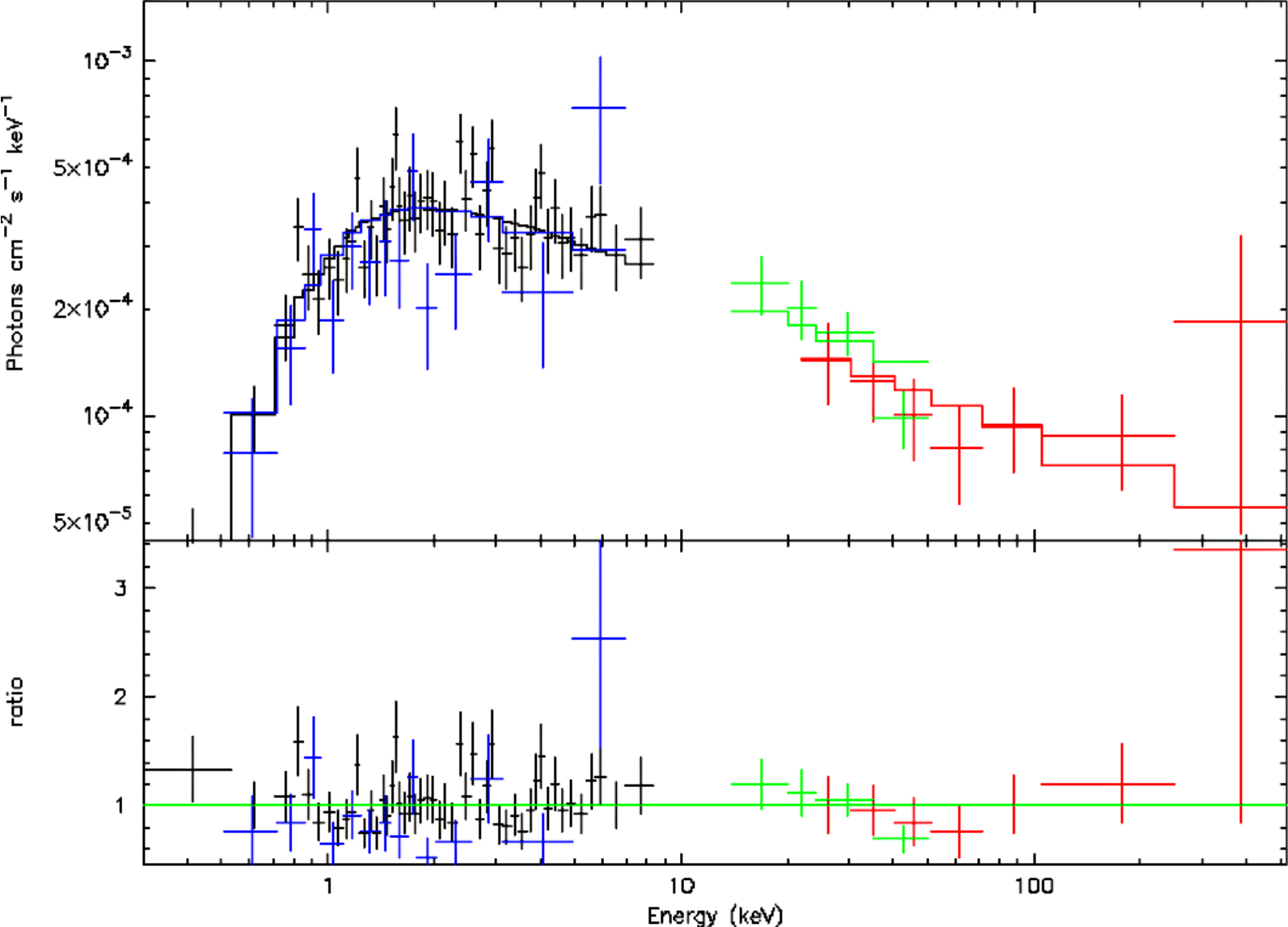}
\caption{Broadband X-ray spectrum of CSRN obtained with {\em eROSITA}~(blue), {\em Swift}\,XRT~(black), {\em Swift}\,BAT~(green), and {\em INTEGRAL} IBIS/ISGRI~(red). The solid line shows the best-fitting absorbed power-law model. The lower panel displays the ratio of the data to the model.}
\label{fig:xrt-bat}
\end{figure}

We summarise the available X-ray observations of CSRN as follows. The {\em Swift}\,XRT data, shown in Fig.\,\ref{fig:xrt-bat}, were obtained using the XRT product generator\footnote{\href{https://www.swift.ac.uk/user\_objects/}{Build Swift-XRT products}}, while the {\em Swift}\,BAT and {\em INTEGRAL}\,IBIS/ISGRI data were extracted using the HEAVENS online tool\footnote{\href{https://www.astro.unige.ch/integral/heavens}{HEAVENS online tool}}. The eRASS1 spectral products were downloaded from the eROSITA--DE DR1 public archive\footnote{\href{https://erosita.mpe.mpg.de/dr1/}{eROSITA--DE DR1}}. The combined X-ray spectrum is well described by an absorbed power-law model \texttt{tbabs$\times$po} fitted in XSPEC, yielding a hydrogen column density of $N_H=3.1\pm0.3 \times 10^{21} \rm{cm}^{-2}$ and a photon index of $\Gamma$\,=\,1.36\,$\pm$\,0.03. These parameters give a good fit with $\chi^2=203.3$ for 203 degrees of freedom. The fluxes measured by each instrument are provided in Table~\ref{tab:index}. 



\begin{table}[htp!]
\caption{Observed X-ray fluxes of CSRN measured with {\em INTEGRAL}\,IBIS/ISGRI, {\em Swift}\,BAT, {\em Swift}\,XRT and {\em eROSITA}~(eRASS1) in their respective energy bands.}
\label{tab:index}
\begin{tabular}{|c | c | c |}
\hline
\makecell{Survey} & \makecell{Flux \\ $[10^{-12}$~erg~cm$^{-2}$~s$^{-1}]$} &  \makecell{Energy Band \\ $[$keV$]$}\\
\hline\hline
{\em INTEGRAL}\,ISGRI & $4.7 \pm 0.3$ / $9.9 \pm 0.3$ & 20-40/40-100\\
{\em Swift}\,BAT & 4.7$\pm$0.3 & 20--40 \\ 
{\em Swift}\,XRT & 3.4 $\pm$ 0.1 & 2--10  \\  
{\em eROSITA} & 3.3$\pm$ 0.1 & 0.2--8 \\
\hline
\end{tabular}
\end{table}

The absorption column density measured from X-ray spectra is
consistent with the total Galactic column density in the direction of \g\ if the CSRN had a distance  $>$\,4 kpc. On the other hand, at a distance of 267\,pc, the measured column density would require additional internal absorption in the system corresponding to the extinction E\,(B--V)\,$\sim 1$ \citep{2009MNRAS.400.2050G}.

\subsection{\erass\ analysis}
\label{sec:coronal}

The optical counterpart of \erass\ is the Gaia~32 star. Gaia~32 has an apparent magnitude of $G$\,$\approx$\,18.45\,mag and lies at a distance of $d$\,$\approx $\,267$\pm$\,9\,pc~\citep{2023AA...674A...1G}. {\em Gaia} DR3 photometry gives a colour of $BP$\,--\,$RP$\,$\approx$\,2.89\,mag, corresponding to a reddening of $E_{BP-RP}$\,$\approx$\,$0.38$\,mag and an effective temperature of $T_\mathrm{eff}$\,$\approx$\,3323\,K. These parameters are consistent with an M3 spectral type. In the Gaia colour-magnitude diagram, the source lies approximately 0.7\,mag below the main sequence. 

In \cite{2024A&A...684A.121F}, the source is flagged as non-coronal based on several properties atypical for stellar coronal emission. Coronal X-ray emitters usually exhibit soft spectra, whereas \erass\ shows relatively hard spectrum with photon index $\Gamma$\,=\,1.36\,$\pm$\,0.03. Another key diagnostic is the X-ray-to-bolometric flux ratio. For coronal emitters, this ratio typically saturates at $\log(F_X/F_\mathrm{bol})$\,$\approx$\,--3 \citep{vil84,wright11} and may increase for individual sources by roughly one order of magnitude due to variability \citep{2024A&A...684A.121F}. For Gaia\,32, however, we derive a much higher value of $\log(F_X/F_\mathrm{bol})$\,=\,$-$0.7 when adopting stellar bolometric corrections. Such a value exceeds those observed even in the most active coronal systems, including RS Canum Venaticorum-type binaries \citep{Dempsey1993}. Additionally, the eRASS1 source exhibits a stable light curve. Taken together, these properties indicate that the observed X-ray emission cannot be explained by coronal activity from an M3 star alone~\citet{2024A&A...684A.121F}. Such a hard spectrum is typical of accreting compact objects or \ac{pwn}, but is unusual for stellar coronal emission.

\subsection{Chance alignment probabilities}
\label{sec:ChanceAlignment}

We treat the hard X-ray detections~({\em INTEGRAL}, {\em Swift\,BAT}, {\em Swift\,XRT} and {\em eROSITA}) and the compact central radio source as physically associated, based on their mutual positional consistency~(Table\,\ref{tab:associations}) and the broad-band spectral agreement~(Sect.~\ref{sec:x-ray}). The only \emph{a priori} questionable associations are the optical/IR objects~({\em Gaia} and 2MASS sources) with the X-ray source, and the probability of finding them projected near the centre of the radio shell.

To quantify these chance-coincidence probabilities, we model the spatial distribution of unrelated sources as a Poisson process and compute the probability\footnote{$p = 1 - e^{(-\lambda\,A)}$, where $\lambda$ is the local surface density of sources and $A$ is the considered search area.} of finding at least one unrelated catalogue object within the given area. Because {\em Gaia} and 2MASS detections are not statistically independent, we use the 2MASS catalogue\footnote{\href{https://irsa.ipac.caltech.edu/cgi-bin/Gator/nph-scan?mission=irsa\&submit=Select\&projshort=2MASS}{2MASS catalogue}} to characterise the population of sources detectable in both the optical and near-IR within 1$^{\circ}$ of \g. After converting the $1\sigma$ positional uncertainty of the eRASS1 source to a 90\% confidence radius, we estimate the probability that an unrelated {\em Gaia}/2MASS source lies within the eRASS1 
localisation region to be $p_{Gaia\bigcap2MASS,eRASS1}=0.0687$. This value should be regarded as a conservative upper limit, as no filtering on magnitude, colour, or stellar type has been applied, and the 2MASS surface density provides only an approximation to the joint {\em Gaia}-2MASS population. 
Similarly, we then calculate the probability for a hard X-ray source to fall within the central 10\% of the radio shell area~(corresponding to a radius of 5.6\arcmin). Using the eROSITA--DE DR1 archive, we obtain a probability of $p_{eRASS1,10\%SNR}=0.0017$. 

Assuming approximate independence between the spatial distribution of eROSITA sources within the inner 10\% of the radio shell and the presence of unrelated {\em Gaia}/2MASS sources within the eROSITA localisation region, the combined probability of such a chance configuration is $p_{total}=0.0012$~(0.12\%). 
This estimate should be regarded as an upper limit to the true chance-alignment probability. A combined probability at the level of $\sim$$10^{-3}$ indicates that a purely chance superposition of the X-ray sources, the optical/near-IR counterpart, and the radio shell is highly unlikely.

\section{On the nature of \g\ and CSRN}
\label{sec:discussion}

CSRN was initially classified as an \ac{lmxb}. By analogy with the small number of high-mass X-ray binaries associated with SNRs, the surrounding radio shell \g\ could in principle be interpreted as the remnant of the SN that formed the compact object \citep[e.g.,][]{2012MNRAS.420L..13H,2021MNRAS.504..326M}. However, the \ac{lmxb} phase is expected to occur long after the SN explosion, on timescales incompatible with the presence of a detectable remnant \citep{Lin2011}. Moreover, the extreme stellar mass ratios required for such systems make the survival of the binary through the SN event unlikely. We therefore consider alternative explanations for the nature of CSRN and its relationship to \g.

\subsection{Nearby low-luminosity SN with a peculiar central source}
\label{subsec:scenario1}

Given the low probability of chance alignment, we consider the radio, X-ray, and optical/near-IR emission to be physically associated, placing the system at the distance of $d$\,$\approx $\,267$\pm$\,9\,pc. 

The morphology of radio shell \g~(Fig.\,\ref{fig:rgb}), together with the absence of extended X-ray emission, suggests a dynamically evolved SNR, similar to several recently detected faint remnants \citep[e.g.,][]{2023AJ....166..149F, 2024PASA...41..112F}. The current shock velocity is therefore likely a few hundred km\,s$^{-1}$, such that freshly accelerated electrons reach only modest maximum energies and are unable to produce detectable non-thermal X-ray emission~(e.g., as observed in RCW\,86~\citep{2011ApJ...741...96W}; see also {\citealt{2006ApJ...648L..33V}). At the same time, the relatively uniform radio morphology does not indicate strong interaction with a dense interstellar cloud, which would produce localised enhancements in radio brightness. The emission suggests that the shock must remain sufficiently strong to sustain the acceleration of radio-emitting electrons. This suggests a shock velocity in the range $\sim$100\,km\,s$^{-1}$\,$\lesssim v_\text{sh}\lesssim$\,1000\,km\,s$^{-1}$. 


Assuming a small ejected mass, \g\ would already be in the Sedov phase; however, for a canonical explosion energy of 10$^{51}$~erg, this would imply an age of only $\sim$100\,yr~(see Table~\ref{tab:age}), which is implausibly young. 
This tension can be resolved if \g\ resulted from a low-energy explosion with kinetic energies well below the canonical core-collapse value. For explosion energies of $10^{47}-10^{48}~$erg, the inferred remnant age increases to several hundred--thousands years while maintaining shock velocities sufficient to produce the observed radio synchrotron emission. Such energies are consistent with weak ccSNe involving neutrino-induced mass loss or ecSNe, both of which require lighter progenitors~($8-10\,M_\odot$). These explosions are expected to impart relatively small natal kicks to the NS, increasing the likelihood of binary survival. These explosion scenarios are explored in more detail in Sect.~\ref{sec:explosionScenario}. 

CSRN has an X-ray luminosity of $L_X\approx3\times10^{31}~$erg\,s$^{-1}$, and a hard spectrum with photon index $\Gamma$\,=\,1.36\,$\pm$\,0.03~(Sect.~\ref{sec:x-ray}). These properties are difficult to reconcile with stellar coronal emission~(Sect.~\ref{sec:coronal}) and lie orders of magnitude below typical luminosities of LMXB \citep[$L_X$\,>\,10$^{34}~$erg\,s$^{-1}$,][]{2023hxga.book..120B}.
Instead, they are consistent with those of spider pulsars, a class of millisecond pulsars~(MSPs) in compact binaries with low-mass companions \citep{2019Galax...7...93H}. In these systems, a relativistic pulsar wind collides with and blows away the outer layers of the optical star, producing high-energy emission 
from an intra-binary shock rather than from accretion. 

We interpret Gaia\,32 as a low-mass companion interacting with a recently formed NS, representing a pre–MSP evolutionary stage. In this scenario, CSRN may eventually evolve into a classical spider pulsar once mass transfer spins up the NS and ablation of the companion becomes efficient. As such, CSRN can be viewed as a candidate baby-spider system, observed shortly after NS formation, possibly in a wider and eccentric orbit, as expected following a natal kick.

\subsubsection{Low-energy explosion scenarios}
\label{sec:explosionScenario}

We explore several plausible explosion mechanisms for \g\ by comparing their characteristic energy scales with the dynamical constraints inferred from the remnant size and ambient density. In Table~\ref{tab:age}, we summarise the Sedov age estimates obtained for several assumed distances, adopting ambient H\,{\sc i} densities from \cite{2006A&A...459..113M}. 
For distances of a few kpc, the derived ages range from a few to several tens of kyr, consistent with an evolved SNR expanding in a low-density medium. In contrast, adopting the nearby distance of the Gaia\,32 counterpart yields an extremely young age of only $\sim$100~yr for a canonical explosion energy. Therefore, reconciling the observed remnant size with a small distance requires a substantially lower explosion energy, motivating the exploration of low-energy explosion scenarios.

\begin{table}[ht!]
\caption{SNR age estimates for various distances assuming a Sedov solution\tablefootnote{~$t_\text{Sedov} = R_\text{SNR}^{5/2}\left(\frac{\alpha\rho}{E}\right)^{1/2}$, where $E=10^{51}\,$erg is the explosion energy, $\rho$ is the ambient density, $R_\text{SNR}$ is the radius of the remnant, and $\alpha$ is a numerical factor of order unity.}. H\,{\sc i} densities are adopted from \cite{2006A&A...459..113M}.}
\label{tab:age}
\hspace{-2mm}
\begin{tabular}{|c|c|c|c|}
\hline
Distance\,[kpc] & Radius\,[pc] & H\,{\sc i} density\,[cm$^{-3}$] & Age\,[kyr] \\
\hline\hline
0.267 & 1.4   & 0.19  & 0.137  \\
3     & 15.5  & 0.10  & 5.4  \\
4     & 20.6  & 0.10  & 11.0  \\
5     & 25.8  & 0.10  & 19.2  \\
\hline
\end{tabular}
\end{table} 

\vspace{-3mm}
\paragraph{\bf (Classical) Novae}
are known to be efficient particle accelerators \citep{2022Sci...376...77H}, with typical explosion energies of $E\approx10^{44}~$erg and a theoretical upper limit of $\sim$$10^{45}~$erg~\citep{2022PASJ...74.1005K}. Although recurrent novae may reach higher cumulative ejecta energies, no nova has been reported in the direction of \g. 
For explosion energies of $10^{44}$–$10^{45}~$erg, the inferred Sedov ages span $\sim$18$-$68~kyr. At these ages, the present-day shock velocities would be only 42\,$-$\,93~km\,s$^{-1}$, far too low to account for the observed non-thermal radio synchrotron emission. We therefore rule out a nova origin for \g.

\vspace{-3mm}
\paragraph{\bf Neutrino mass-loss explosions} 
Weak core-collapse explosions driven by neutrino-induced mass loss have been proposed as a possible outcome of stellar core collapse when the explosion energy is insufficient to launch a strong outgoing shock. Numerical simulations by~\cite{2013ApJ...769..109L} show that even in the absence of an outward propagating shock, the hydrodynamic response to abrupt neutrino mass loss from the collapsing core can generate a weak explosion with characteristic energies of order $E$\,$\sim$\,$10^{47}~$erg. 
Adopting such an explosion energy for \g\ yields an inferred remnant age of $\sim$1.8~kyr and a present-day shock velocity of $\sim$400~km\,s$^{-1}$, consistent with efficient particle acceleration and the observed radio synchrotron emission.

\vspace{-3mm}
\paragraph{\bf ILRT theoretical expectations}
ILRT have been proposed as the observational outcome of low-energy ecSNe~\citep{2019NatAs...3..676P}. Early theoretical models predict explosion energies of order $E$\,$\sim$\,$10^{50}~$erg for ecSNe~\citep{2006A&A...450..345K}. However, adopting this energy for \g\ would place the remnant in the free-expansion phase, implying an age of only $\sim$$100$~yr and an unrealistically high shock velocity, inconsistent with the observed radio shell.

\vspace{-3mm}
\paragraph{\bf ILRT observational constraints}
Recent observational studies of ecSNe from progenitors of $8$–$10\,M_\odot$ mass range suggest substantially lower explosion energies of order $E$\,$\sim$\,$10^{48}~$erg~\citep{2020A&A...639A.103S}. Adopting such an energy for \g\ yields a physically plausible solution, with an age of $\sim$$600$~yr and a shock velocity of $\sim$700~km\,s$^{-1}$, consistent with efficient particle acceleration and the observed non-thermal radio emission.

\subsection{Distant SNR hosting a young rotation-powered pulsar}
\label{subsec:scenario2}

To test whether a conventional interpretation can explain the observed properties, we also consider a scenario in which CSRN represents a \ac{pwn} located at the centre of \g. In this case, the apparent spatial coincidence between the X-ray and optical/near-IR sources is due to chance alignment.
The hard X-ray spectrum of CSRN is consistent with that observed for young rotation-powered pulsars~(RPPs) and their \ac{pwn}e \citep{2008AIPC..983..171K}. The radio spectral index flattens toward the core and is compatible with synchrotron emission from a \ac{pwn} \citep{2017ASSL..446....1K}. In addition, CSRN exhibits an X-ray to radio luminosity ratio of $L_X/L_R$\,$\approx$ $9000$, comparable to the highest values observed for young \ac{pwn}e.

The morphology and dynamical properties of the radio shell indicate an evolved SNR, with an age of $\sim$10\,kyr at a distance of a few kpc assuming a canonical explosion energy of $10^{51}$\,erg~(Table~\ref{tab:age}). Adopting a fiducial distance of 4\,kpc yields an X-ray luminosity of $L_X$\,$\sim$\,$10^{34}\,$erg\,s$^{-1}$ and a 1.4\,GHz radio luminosity of $L_R$\,$\sim$\,$10^{30}\,$erg\,s$^{-1}$ for CSRN. 
While the X-ray luminosity is moderate, the radio luminosity places CSRN at the extreme low end of young \ac{pwn}e. Thus, the large luminosity ratio arises from suppressed radio emission rather than unusually strong X-ray output. Such behaviour is difficult to explain with standard evolutionary models, in which young \ac{pwn}e are also radio bright~\citep{2017ASSL..446....1K}.

Assuming an optimistic X-ray efficiency~($\eta$\,=\,$0.1$), the inferred luminosity implies a pulsar spin-down power of at least $10^{35}\,$erg\,s$^{-1}$\,\citep{2011ApJ...727..131V}. Pulsars with comparable energetics typically produce detectable $\gamma$-ray emission, particularly at off-plane locations where background contamination is low~\citep{2023ApJ...958..191S}. However, the closest \textit{Fermi}--LAT source is associated with AGN\,MGPS\,J111715--533816 and $\sim$$1^\circ$\,offset from CSRN~\citep{2023arXiv230712546B}. We also conducted a 
pulsar search observation, but no pulsations were detected~(Appendix~\ref{parkes-data}). Hence, even if an isolated NS scenario appears less exotic, the unusual properties of CSRN indicate that it is special.

\section{Summary and conclusions}
\label{sec:conclusion}

We report the discovery of \g, a low surface brightness radio shell identified in ASKAP--EMU data~(Fig.\,\ref{fig:rgb}). Its morphology and non-thermal spectrum strongly support an SNR interpretation. 
A compact radio nebula at its centre hosts the soft $\gamma$-ray source \igr. The $\gamma$-ray source is spatially coincident with a hard X-ray source previously classified as an LMXB, with an M3-type stellar optical counterpart. However, adopting its Gaia DR3 distance implies the X-ray luminosity is far too low for a typical LMXB. 

Based on the currently available data, \g's properties favour a low-energy explosion from an intermediate mass progenitor in which a young NS remains bound to its stellar companion. If confirmed, \g\ would represent the first Galactic remnant of an ILRT hosting a binary system, providing a unique link between such transients, NS formation and the early evolution of spider pulsars.
A more conventional alternative in which the optical star is a chance alignment and IGR J11187--543 is an isolated NS embedded in a distant evolved SNR cannot be excluded. However, it would require an unusually radio-faint \ac{pwn} with a high X-ray to radio luminosity ratio.

Further multiwavelength follow-up observations are essential to distinguish between these scenarios and to establish the true nature of this intriguing system.

\begin{acknowledgements} 
The authors are grateful to Dr.~Jes\'us Maiz Apell\'aniz for his insights and advice on the properties of \gaia\ as inferred from the {\em Gaia} catalogues. 

This scientific work uses data obtained from Inyarrimanha Ilgari Bundara, the CSIRO Murchison Radio-astronomy Observatory. We acknowledge the Wajarri Yamaji People as the Traditional Owners and native title holders of the Observatory site. CSIRO’s ASKAP radio telescope is part of the Australia Telescope National Facility~(ATNF)(https://ror.org/05qajvd42). Operation of ASKAP is funded by the Australian Government with support from the National Collaborative Research Infrastructure Strategy. ASKAP uses the resources of the Pawsey Supercomputing Research Centre. Establishment of ASKAP, Inyarrimanha Ilgari Bundara, the CSIRO Murchison Radio-astronomy Observatory and the Pawsey Supercomputing Research Centre are initiatives of the Australian Government, with support from the Government of Western Australia and the Science and Industry Endowment Fund. 
Murriyang, CSIRO’s Parkes radio telescope, is part of the ATNF. We acknowledge the Wiradjuri people as the Traditional Owners of the Observatory site.

This work uses observations obtained at the Southern Astrophysical Research~(SOAR) telescope, which is a joint project of the Minist\'{e}rio da Ci\^{e}ncia, Tecnologia e Inova\c{c}\~{o}es~(MCTI/LNA) do Brasil, the US National Science Foundation’s NOIRLab, the University of North Carolina at Chapel Hill~(UNC), and Michigan State University~(MSU).

SL, MDF, and GR acknowledge the Australian Research Council funding through grant DP200100784.

ZG and JO are supported by the China-Chile Joint Research Fund~(CCJRF No.2301), the Chinese Academy of Sciences South America Center for Astronomy~(CASSACA, Key Research Project E52H540301), and ANID, Millennium Science Initiative~(AIM23-001).

BvS acknowledges support by the National Research Foundation of South Africa~(grant number 119430). 


\end{acknowledgements}
\bibliographystyle{aa}
\bibliography{G289}
%
%

\appendix 

\section{ASKAP--EMU observations and processing}
\label{sec:askap-obs}

The Australian Square Kilometre Array Pathfinder~\citep[ASKAP; ][]{2021PASA...38....9H} is a radio interferometer located at Inyarrimanha Ilgari Bundara, the CSIRO Murchison Radio-astronomy Observatory in Western Australia. The array consists of 36 fixed 12~m antennas, with baselines ranging from 22~m to 6.4~km. Each antenna is equipped with a phased array feed~(PAF) at the primary focus, forming 36~beams simultaneously and providing a large field of view of $\sim$$25-30$~deg$^2$. This design makes ASKAP a powerful instrument for conducting large-area radio surveys.

We used archival data from ASKAP's deep radio-continuum survey, the Evolutionary Map of the Universe~\cite[EMU\footnote{\href{https://emu-survey.org/}{EMU Project page}};][]{2021PASA...38...46N,2025PASA...42...71H}, which aims to produce a comprehensive atlas of the southern sky and is currently expected to be completed in 2028. EMU observations are centred at 944~MHz with a bandwidth of 288\,$\times$\,1~MHz channels. Standard observations have a full 10~h track, achieving a typical sensitivity of 25$-$30~$\mu$Jy\,beam$^{-1}$. Data processing is performed using the ASKAPsoft pipeline~\citep{Guzman_Askapsoft}. The final data products are available through the CSIRO ASKAP Science Data Archive~(CASDA\footnote{\href{https://data.csiro.au/domain/casdaObservation}{CSIRO ASKAP Science Data Archive~(CASDA)}\label{casda}}), under project code AS201.

\section{EMU data products}
\label{sec:emu-data}

The EMU survey produces three main types of full-band images for each field: ``conv'', ``raw'', and ``highres''. The standard images, identified by the ``conv'' suffix in the filename, have a uniform resolution of $15^{\prime\prime}$, ensuring consistent flux density measurements across the survey. The ``raw'' images use the native beam, resulting in higher resolution~($\sim$11$^{\prime\prime}$--13$^{\prime\prime}$), but variations in beam size across a field make flux density measurements less reliable. The ``highres'' images are produced using uniform weighting and achieve even higher resolution~($\sim$7$^{\prime\prime}$--9$^{\prime\prime}$). While these images enhance fine structure in bright emission, they suffer from reduced sensitivity to diffuse emission and are therefore unsuitable for quantitative flux-density analysis.

For this work, we used the recommended ``conv'' image for Fig.\,\ref{fig:rgb} and for the quantitative analysis of \g. We also examined the corresponding ``highres'' image, which has a significantly higher resolution of $\sim$7.1$^{\prime\prime}\times$7.9$^{\prime\prime}$. Both the ``conv'' and ``highres'' images are shown in Fig.\,\ref{fig:conv-highres}. As expected, flux inconsistencies arise from reduced sensitivity of the ``highres'' images to diffuse emission. Nevertheless, the ``highres'' data reveals a point source in the northeast of the central nebula at ~(R.A.,\,Dec)$_\text{J2000}$~= 11:18:21.27, --54:37:29.88. The peak, \askap, positionally coincides with the Gaia\,32, \erass, and \mass\ sources. In addition, the ``highres'' contours highlight a secondary peak toward the southwest; however, this feature does not resemble a compact source and is likely caused by flux fluctuations inherent to this data product.

\section{Spectral analysis methods}
\label{sec:spectral}

To constrain the spectral shape and derive the spectral indices, flux density measurements at multiple widely separated frequencies are typically required. However, flux measurements for \g\ were only available from the EMU survey. In this case, we used two in-band methods: sub-band imaging and the Taylor term technique. Both methods are limited to high signal-to-noise, so we restricted our analysis to the central region and the brightest northeastern filamentary structures.

\vspace{-3mm}
\paragraph{\bf Sub-band imaging} We used image cubes with a total bandwidth of 288\,MHz, covering the frequency range from 800 to 1088\,MHz at 1\,MHz resolution. The calibrated visibilities from beams that cover \g\ were split into four 72\,MHz spectral windows, and reprocessed using ASKAPsoft \citep{Guzman_Askapsoft}. The resulting sub-band images were primary beam-corrected, mosaicked, and convolved to a common resolution of $18''\times18''$. These final images are centred to reference frequencies of 835, 908, 980, and 1051\,MHz.

Following the method described in \cite{2023MNRAS.524.1396B}, we measured the integrated flux density of the central region of \g\ using \textsc{Polygon\_Flux} software \citep{Flux_GLEAM}. Flux densities were extracted from three polygonal regions defined by cyan contour levels shown in Fig.\,\ref{fig:conv-highres} and fitted with a simple power law to derive radio spectral indices\footnote{\label{noteAlpha}$S\propto\nu^{\alpha}$, where $S$ is the flux density, $\nu$ is the frequency, and $\alpha$ is the spectral index.}. The resulting spectral indices are $-0.51\pm0.26$, $-0.38\pm0.18$ and $-0.28\pm0.10$ for the outer, middle and inner polygons, respectively. The derived values are summarised in Table~\ref{tab:flux_central}. For the brightest filament segments to the northeast, we derive spectral indices of $-0.76 \pm 0.56$ for the outer filament and $-0.56 \pm 0.32$ for the inner filament.

\paragraph{\bf Taylor term technique} 
ASKAPsoft models the wide-band sky brightness distribution as a Taylor series expansion in frequency, producing a set of Taylor-coefficient images \citep{2021PASA...38....9H,2021PASA...38...46N}. We use the first two terms: the zeroth-order term \texttt{T0}, representing the total intensity integrated over the observing band, and the first-order term \texttt{T1}, which describes the frequency dependence of the intensity.  Assuming a power-law spectrum\footref{noteAlpha}, the spectral index is computed as $\alpha~=~\texttt{T1}/\texttt{T0}$. 

We generated the Taylor term images from the same calibrated visibilities used in the sub-band imaging, and convolved to a common resolution of $18'' \times 18''$ to allow direct comparison. To suppress noise artefacts in low signal-to-noise regions, we applied a $3\sigma$ flux density mask based on \texttt{T0} image. 

Using the same polygon regions as in the previous method, we derive mean spectral indices of --0.51\,$\pm$\,0.20, --0.34\,$\pm$\,0.10, and $-0.22 \pm 0.10$ for the outer, middle, and inner regions of the central source~(Table~\ref{tab:flux_central}), respectively, and $-0.71 \pm 0.45$ and $-0.59 \pm 0.40$ for the bright northeastern, outer and inner, filaments. 

\begin{table}[hp!]
\centering
\caption{Integrated flux densities and spectral indices of the central nebula, measured within three regions defined by different contour levels~(shown as cyan contours in Fig.\,\ref{fig:conv-highres}). Measurements are based on the full- and four sub-band ASKAP--EMU images.}
\label{tab:flux_central}
\begin{tabular}{|c|c|c|c|c|}
\hline
\makecell{Frequency\\ $[$MHz$]$} & \makecell{Bandwidth \\ $[$MHz$]$} &\makecell{Central \\ outer \\ S$_{\nu}$\,$[$mJy$]$} & \makecell{Central \\ middle \\ S$_{\nu}$\,$[$mJy$]$}  & \makecell{Central \\ inner \\ S$_{\nu}$\,$[$mJy$]$}\\
\hline\hline
 944  & 288  & 31.08 & 19.98 &  9.11 \\
\hline
 835  & 72  & 32.67  & 20.66 & 9.42 \\
 908  & 72  & 31.28  & 19.99 & 9.20 \\ 
 980  & 72  & 30.47  & 19.74 & 9.00 \\
1051  & 72  & 28.94  & 18.80 & 8.83 \\
\hline
$\alpha_\text{sub-imaging}$ &  & --0.51$\pm$0.26 & --0.38$\pm$0.18 & --0.28$\pm$0.10\\
$\alpha_\text{Taylor}$ &  & --0.51$\pm$0.20 & --0.34$\pm$0.20 & --0.22$\pm$0.20\\
\hline
\end{tabular}
\end{table}

\section{Optical observations}
\label{sec:optical} 

Optical spectroscopy of the Gaia\,32 was obtained with the 4.1~m Southern Astrophysical Research~(SOAR) telescope located on Cerro Pachón in Chile, using the Goodman longslit spectrograph~\citep{Clemens2004}. The observations were carried out in two epochs on 21~June and 24~July 2025~(PI: I. El Mellah). During each epoch, we acquired two spectra using the 400\,M2 grism mounted on the RED camera with a 1\arcsec\ slit and 2$\times$2 binning, providing wavelength coverage from 5000 to 9050~\AA, and a 3-pixel spectral resolution of 13\,\AA. Each exposure had an integration time of 600~s. 

During the second observing epoch, we used two different slit orientations to minimise contamination from nearby sources. The data were reduced using the Goodman spectroscopic pipeline, including standard flat-field and ARC calibrations. One-dimensional spectra were extracted from the reduced two-dimensional frames produced by the pipeline and combined within each epoch. Flux calibration was performed using a spectrophotometric standard star observed during the same night, with atmospheric extinction corrections derived from the accompanying $g$- and $r$-band acquisition images.

%

\begin{figure}[htp!]
\includegraphics[trim =35 15 15 25 , width=1\columnwidth]{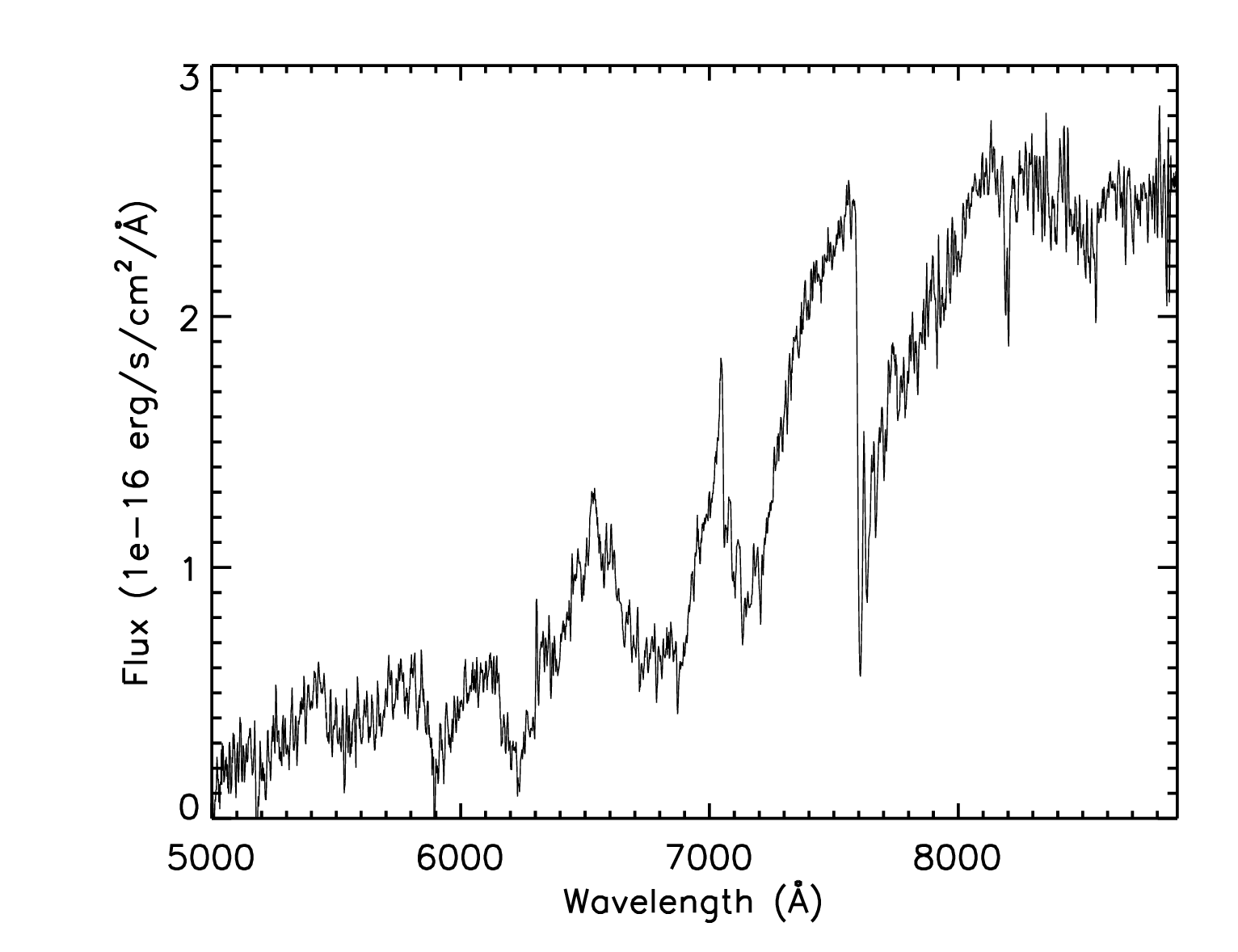}
\caption{SOAR Optical spectrum of Gaia\,32 with a total integration time of 3600\,s.}
\label{fig:optical}
\end{figure}

The combined spectrum shows clear photospheric features of a cool star, including prominent molecular absorption bands, such as TiO and VO. By comparison with archival spectral models~\citep[e.g., BT-Settl;][]{Allard2011}, we estimate a spectral type of M3 with an effective temperature of $T_{\rm eff}$\,$\approx$\,3200\,--\,3300\,K and a line of sight extinction of A$_V$\,$\approx$\,0.5\,--\,0.6\,mag. The spectral classification is primarily based on the flux ratios outside and inside the TiO absorption bands. The spectrum is presented in Fig.\,\ref{fig:optical}. 

\begin{figure}
\includegraphics[width=\columnwidth]{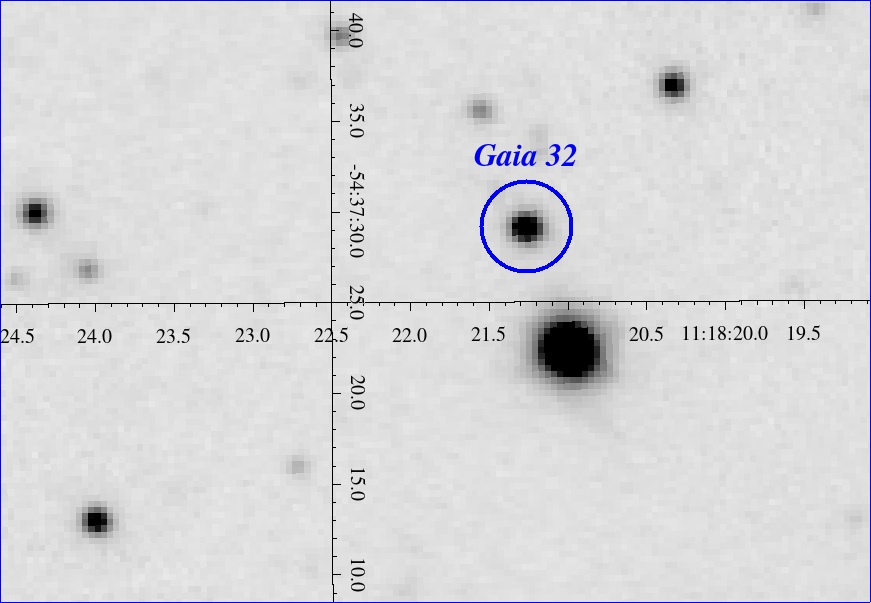}
\caption{Near-IR J-band image of the field around \igr, retrieved from the ESO Science Archive. The image was obtained with the VIRCAM instrument mounted on the ESO VISTA telescope on 7 December 2013 under $1\arcsec$ seeing conditions, with an astrometric rms $0.05\arcsec$. The blue circle~($r=2.5\arcsec$) is centred on the Gaia\,32~(Table\,\ref{tab:associations}). The image is shown on the linear scale.}
\label{fig:ir}
\end{figure}

Optical photometry of Gaia\,32 was obtained on 22 and 29~June 2025, and 11~July 2025, using the SHOC cameras~\citep{2013PASP..125..976C} mounted on the 1.9~m and 1.0~m telescopes at the South African Astronomical Observatory~(SAAO). Exposures were triggered by a GPS-controlled timing system to ensure precise time stamps. Observations were taken using an R~(22~June), $r^\prime$~(29~June), and clear~(11~July) filters, with each observing run lasting approximately 3~h. Differential photometry was performed for the target using nearby comparison stars within the same field of view to search for variability, following standard aperture-photometry procedures. As with the spectroscopic observations, the photometric analysis was affected by contamination from a nearby bright star~(Fig.\,\ref{fig:ir}), which limited the sensitivity to small-amplitude variability. 
\vspace{-2px}

\section{Pulsar search with Murriyang}
\label{parkes-data}

Following the discovery, we carried out a deep pulsar search observation of the central region of \g\ using Murriyang, CSIRO's Parkes radio telescope, with Ultra-Wideband Low~(UWL) receiver in conjunction with the Medusa backend~\citep{hobs2020}. The observation was obtained on 20 October 2023 under project code P1238~(PI:\,A.\,Ahmad). The total integration time was 12,545\,s, recording total-intensity data only. The data were acquired in pulsar search mode with 2-bit sampling at a time resolution of 64\,$\mu$s and frequency channels of width 0.125\,MHz, resulting in 26,624 channels covering the frequency range 704--4032\,MHz. 

We segmented the UWL data into three frequency sub-bands, with a mid-band ranging from 1344--2112\,MHz~(centred at 1728\,MHz), and a high-band ranging from 2112--3520\,MHz~(centred at 2816\,MHz). The mid-band provides good sensitivity to low \ac{dm} pulsars with steep spectra, while the higher-frequency band reduces the effects of \ac{dm} smearing and scattering, improving sensitivity to high-\ac{dm} pulsars. The lowest part of the UWL band~(below 1344\,MHz) was heavily contaminated by \ac{rfi} and was therefore excluded from the analysis. 

A periodicity search was performed using the pulsar search software 
package \texttt{PRESTO}\footnote{\href{https://www.proquest.com/docview/304699474}{\texttt{PRESTO}: a search technique for binary pulsars}} \citep{2002AJ....124.1788Ransom}, covering a \ac{dm} range of 0--2000~pc\,cm$^{-3}$. \ac{rfi} was mitigated using the \textit{rfifind} routine. To account for potential orbital modulation, we searched for signals allowing them to drift by up to $\pm200/n_{\rm h}$ bins in the Fourier domain by setting $z_{\rm max}=200$, where $n_{h}$ is the highest detected harmonic~(up to eight harmonics were summed). Candidates with a signal-to-noise ratio greater than 8 were folded and inspected visually. No pulsations were detected in either frequency band. The data are publicly available through the Parkes Pulsar Data Archive\footnote{\href{https://data.csiro.au/domain/atnf}{Parkes Pulsar Data Archive~(PPDA)}}.

\label{LastPage}
\end{document}